\documentstyle[12pt,aaspp4]{article}
\begin{document}
\parindent=1.0cm

\title{THE DETECTION OF BRIGHT ASYMPTOTIC GIANT BRANCH STARS IN THE NEARBY ELLIPTICAL GALAXY MAFFEI 1}

\author{T. J. Davidge \altaffilmark{1}}

\affil{Canadian Gemini Office, Herzberg Institute of Astrophysics,
\\National Research Council of Canada, 5071 W. Saanich Road, \\ Victoria, 
British Columbia, Canada V9E 2E7 \\ {\it email: tim.davidge@nrc.ca}}

\author{Sidney van den Bergh}

\affil{Dominion Astrophysical Observatory, Herzberg Institute of Astrophysics,
\\National Research Council of Canada, 5071 W. Saanich Road, \\ Victoria, 
British Columbia, Canada V9E 2E7 \\ {\it email: sidney.vandenbergh@nrc.ca}}

\altaffiltext{1}{Visiting Astronomer, Canada-France-Hawaii Telescope, which is 
operated by the National Research Council of Canada, the Centre National de le 
Recherche Scientifique, and the University of Hawaii.}

\begin{abstract}

	We have used the adaptive optics system on the Canada-France-Hawaii 
Telescope to study a field 6 arcmin from the center of the heavily 
obscured elliptical galaxy Maffei 1. Our near diffraction-limited $H$ and 
$K'$ images reveal an excess population of objects with respect to a background 
field, and we conclude that the asymptotic giant branch tip (AGB-tip) in Maffei 
1 occurs at $K = 20 \pm 0.25$. Assuming that stars at the AGB-tip in Maffei 1 
have the same intrinsic luminosity as those in the bulge of M31, then the 
actual distance modulus of Maffei 1 is $28.2 \pm 0.3$ if A$_V = 5.1$, which 
corresponds to a distance of $4.4^{+0.6}_{-0.5}$ Mpc. This is in excellent 
agreement with the distance estimated from surface brightness flucuations in 
the infrared, and confirms that Maffei 1 is too distant to significantly affect 
the dynamics of the Local Group.

\end{abstract}

\keywords{galaxies: individual (Maffei 1) -- galaxies: distances and redshifts}

\section{INTRODUCTION}

	The elliptical galaxy Maffei 1 (Maffei 1968) 
is one of the dominant members of a group that 
includes the spiral galaxy IC 342 (van den Bergh 1971). This group of galaxies 
contains some of the largest extragalactic objects in the northern sky (Buta \& 
McCall 1999), indicating that Maffei 1 and its companions are relatively close. 
In fact, with the possible exception of NGC 5128, which has a distance of 3.9 
Mpc (Harris, Harris, \& Poole 1999), Maffei 1 may be the nearest giant 
elliptical galaxy and, despite being located at low Galactic latitudes, is thus 
an important benchmark for studies of these systems. Moreover, if Maffei 1 is 
as close as 2 Mpc (Buta \& McCall 1983), then it could significantly 
affect the dynamics of the Local Group, which could then no longer be 
considered as a simple two-body system (Valtonen et al. 1993; Zheng, Valtonen, 
\& Byrd 1991). This has potential cosmological implications, as Local Group 
timing arguments provide a direct geometrical constraint on the mass of the 
Galaxy and M31 (e.g. Kahn \& Woltjer 1959) and, by extension, the dark matter 
content of galaxies.

	Buta \& McCall (1999) list radial velocities for 13 members of the 
IC 342/Maffei group, and the mean is $<v_0> = +30 \pm 25$ km sec$^{-1}$ with a 
dispersion $89 \pm 25$ km sec$^{-1}$. Using 
equation 6 of Courteau \& van den Bergh (1999), the mean velocity of this group 
relative to the Local Group center is then $+279$ km sec$^{-1}$, which 
significantly exceeds the velocity dispersion of the Local 
Group ($61 \pm 8$ km sec$^{-1}$; Courteau \& van den Bergh 1999). Thus, the 
radial velocity of the IC 342/Maffei group is consistent with these objects 
being well outside the Local Group. However, there is a sizeable spread in the 
distances that have been estimated for Maffei 1 and its companions. 
Using the Faber-Jackson relation, Buta \& McCall (1983) concluded 
that the distance to Maffei 1 is $2.1_{-0.8}^{+1.3}$ Mpc. More recently, 
Luppino \& Tonry (1993) obtained a distance of 4.2 $\pm 0.5$ Mpc using surface 
brightness flucuations in $K$, and $4.2 \pm 1.1$ Mpc based on the 
D$_n - \sigma$ relation. The distances to other members of the IC 
342/Maffei group have been estimated by McCall (1989), Krismer, Tully, \& Gioia 
(1995), Karachentsev et al. (1997) and Ivanov et al. (1999), and the results 
fall between 1.7 and 5.3 Mpc. While this large range in distances 
may call into question the reality of an IC 342/Maffei 
group, many of the distance estimates have large 
uncertainties, due in part to the significant amount of reddening towards 
these objects. The radial velocity dispersion of the 
IC 342/Maffei group deduced from the Buta \& McCall (1999) data, quoted in the 
opening sentence of this paragraph, is not significantly different from that 
of the Local Group derived by Courteau \& van den Bergh (1999), and this 
would likely not be the case if the galaxies were spread out over 3.5 Mpc.

	The large range of distances measured, and inferred, for Maffei 1 and 
its companions underscores the need to obtain additional distance estimates. 
Direct constraints on the distance to Maffei 1 can be obtained from the 
photometric properties of resolved stars in this galaxy. A complicating factor 
is that Maffei 1 is viewed at low Galactic latitudes, and efforts to probe the 
stellar content of this galaxy must deal with contamination from stars in the 
Galactic disk and extinction by interstellar dust. The extinction towards 
Maffei 1 is considerable, with A$_V = 5.1 \pm 0.2$ mag (Buta \& McCall 1983), 
and so efforts to study the resolved stellar content of this galaxy are 
restricted to the infrared. If the distance to Maffei 1 is not 
greater than 10 Mpc, and this galaxy has an AGB content similar 
to that in the bulge of M31, which peaks at $K = 15.6 \pm 0.1$ (Davidge 
2001a), then the brightest stars should be detectable in the near-infrared 
using a 4 meter telescope equipped with an adaptive optics (AO) system.
In the current paper, we discuss a preliminary reconnaisance of the 
stellar content of a field 6 arcmin from the center of 
Maffei 1 using deep $H$ and $K'$ images with angular resolutions approaching 
the diffraction limit of the 3.6 metre Canada-France-Hawaii Telescope (CFHT).

\section{OBSERVATIONS AND REDUCTIONS}

	The data were recorded with the CFHT AO Bonette (Rigaut 
et al. 1998) and KIR imager during the nights of UT 2000 September 10 
and 11. KIR contains a $1024 \times 1024$ Hg:Cd:Te array with 
0.034 arcsec pixels, and thus images a $34 \times 34$ arcsec field.

	Two fields were observed through $H$ and $K'$ (Wainscoat \& Cowie 1992) 
filters. One field, centered on the AO reference star GSC 03699-00675 
($\alpha = 02^{h} 36^{m} 05^{s}$, $\delta = +59^{o} 34^{'} 29^{''}$ J2000) 
samples Maffei 1 at a projected distance of 6 arcmin from the galaxy center. 
This field was selected based on (1) the surface brightness of Maffei 1, which 
the measurements given by Buta \& McCall (1999) suggested was sufficiently low 
at this galactocentric radius to permit individual stars on the upper magnitude 
of the RGB to be resolved with the AO system during median seeing conditions, 
and (2) the availablity of an AO reference star. This field will be referred to 
as the `galaxy' field for the remainder of the paper. The second field, which 
will be referred to as the `background' field, is located 25 arcmin from 
Maffei 1, and is centered on the AO reference star GSC 03699 - 01721 
($\alpha = 02^{h} 36^{m} 47^{s}$, $\delta = +59^{o} 09^{'} 06^{''}$ J2000).

	The data were obtained in sets of five 60 sec exposures, which were 
recorded at each corner of a $0.5 \times 0.5$ arcsec square dither pattern. The 
dither sequence was repeated to increase signal from faint sources.
The total exposure time for the galaxy field was 160 minutes per 
filter, while for the background field the total exposure time was 60 
minutes per filter.

	The data were reduced using the procedures discussed by Davidge \& 
Courteau (1999), and the final images have FWHM = 0.14 arcsec ($H$) and 0.15 
arcsec ($K'$).

\section{RESULTS}

	The brightnesses of individual stars were measured with the 
point-spread-function (PSF) fitting routine ALLSTAR (Stetson \& Harris 1988), 
and the instrumental measurements were transformed into the standard system 
using coefficients derived from observations of faint UKIRT standard stars 
(Casali \& Hawarden 1992). A single PSF was constructed for each image
using tasks in DAOPHOT (Stetson 1987); while anisoplanicity 
causes the PSF to vary with distance from the guide star, past 
experience with the CFHT AO system indicates that these variations 
do not dominate the uncertainties in the photometry 
over the KIR field (Davidge \& Courteau 1999; Davidge 2001b). Speckle noise 
hinders the detection of faint sources near the relatively bright AO guide 
stars (Racine et al. 1999), and so the area within 3 arcsec of the guide 
star in each field was excluded from the photometric analysis.

	Artificial star experiments were used to 
estimate completeness and assess systematic effects in the photometry. 
Artificial stars with $H-K$ colors similar to those of detected 
objects, and in numbers small enough to prevent artificially increasing 
the degree of crowding, were added to the images of each field. 
Photometry of the artificial stars indicate that (1) systematic errors in 
the measured brightnesses exceed 0.1 mag when $K \geq 21.5$, and (2) 
the completeness fractions in the galaxy and background fields are similar 
when $K \leq 21.5$; hence, $K = 21.5$ is adopted as the faint limit for these 
data. While the galaxy field might be expected to go slightly deeper than the 
background field because of the difference in integration times, 
this is not the case because of the greater degree of crowding among faint 
objects in the galaxy field.

	Massive elliptical galaxies have sizeable, centrally-concentrated 
globular cluster populations. Buta \& McCall (private communication) have 
detected resolved objects in WFPC2 images centered on the core of Maffei 1, 
and they suggest that these are globular clusters. We have not detected 
similar objects in our galaxy field; this null detection is perhaps not 
surprising given (1) the modest field size, and (2) that we observed 
the outer regions of the galaxy, where the density of clusters is low.

	The $(K, H-K)$ color-magnitude diagrams (CMDs) of the galaxy and 
background fields are compared in Figure 1. Main sequence stars in the Galactic 
disk form a locus with $H-K$ between 0 and 0.4 that dominates the bright 
portions of both CMDs. The main sequence locus has similar colors in both 
fields, indicating that the reddenings along the two sight lines are not 
vastly different.

	The $(K, H-K)$ CMD of the galaxy field contains a population 
of red objects with $K > 20.5$ that is not seen in the background 
field. The statistical significance of these faint stars can be 
investigated by comparing the $K$ luminosity functions (LFs) of the galaxy and 
background fields, and this is done in Figure 2. The LFs in this figure were 
constructed from sources detected in both filters. The LFs of both fields show 
qualitatively similar behaviour at the bright end, rising slowly from $K = 15$ 
to $K = 20$; however, at $K = 20.5$ the LF of the galaxy field 
departs from this trend. The difference 
between the LFs of the galaxy and background fields shows an 
excess at $K = 20.5$ that is significant at the $2.3 \sigma$ level; at 
fainter levels the statistical significance of the excess population becomes 
even greater.

	It is evident from Figure 1 that the galaxy field contains more 
Galactic disk stars than the background field, and the ratio of objects with 
$K$ between 15 and 20 in these fields is $33/17 = 1.9 \pm 0.6$; this difference 
could be due to stochastic variations in star counts, 
clustering, and/or differential absorption. To investigate the extent to which 
the difference in foreground star counts affects the statistical significance 
of any faint population in the galaxy field, the background field LF 
was scaled up by a factor of 1.9, and the difference between the galaxy 
field and scaled background field LFs is shown in the bottom panel of 
Figure 2. Even after scaling the background field LF, the galaxy 
field LF still contains an excess number of objects when $K > 20.5$, 
indicating that the faint population detected in the galaxy field is not an 
artifact of field-to-field variations in foreground star counts. We thus 
conclude that the faint stars seen in the galaxy field belong to Maffei 1.

\section{DISCUSSION}

	The distance to Maffei 1 can be estimated to an internal accuracy of 
$\pm 0.25$ mag based on the star counts in the 
galaxy field. To demonstrate this, assume A$_V = 
5.1$ (Buta \& McCall 1983) and a preliminary distance of 4.2 Mpc (Luppino \& 
Tonry 1993). Since the surface brightness of Maffei 1 in the galaxy 
field is $\mu_V = 27$ mag arcsec$^{-2}$ based on the Buta \& 
McCall (1999) light profile, the total integrated brightness 
in this field is M$_V = -13.8$. Davidge (2001a) discusses observations
of a field in the bulge of M31 with almost the same integrated intrinsic 
brightness as the Maffei 1 galaxy field, and found 4 stars within 0.5 mag of 
the AGB-tip. The number density of stars climbs rapidly with decreasing 
brightness in the M31 bulge field, and there are 58 stars in the next 0.5 mag 
interval. If the stellar content of the Maffei 1 galaxy field is like that in 
the M31 bulge then (1) the number of stars within 0.5 mag of the AGB-tip will 
be modest, to the point where they may not be detectable above the statistical 
noise in the background counts, and (2) there will be a rapid increase in the 
stellar density in the next 0.5 mag interval. Thus, while it might be 
straight-forward to detect an AGB component given the steeply rising LF within 
1 mag in $K$ of the AGB-tip, small number statistics in the top 0.5 mag mean 
that the AGB-tip can likely not be identified to better than $\pm 0.25$ mag 
in fields at moderately large distances from the galaxy center. 

	After correcting for incompleteness and subtracting background star 
counts, we find that there are $35 \pm 15$ stars with $K$ between 20.25 and 
20.75 in the galaxy field, suggesting that this brightness interval samples the 
AGB roughly 0.5 mag below the AGB-tip. We thus consider $K = 20.25$ (i.e. the 
bright edge of the $K = 20.5$ bin) to be the faint limit of the AGB-tip in 
Maffei 1, while the AGB-tip will likely not be brighter than $K = 20.25 - 0.5 = 
19.75$; hence, we adopt $K = 20.0 \pm 0.25$ as the brightness of the AGB-tip.

	Davidge (2001a) finds that the brightest stars in M32, the bulge of 
M31, and the Galactic bulge have similar infrared brightnesses, with M$_K = 
-8.9 \pm 0.1$, suggesting that the AGB-tip can be used as a standard candle to 
estimate the distance to Maffei 1. We estimate the distance to Maffei 1 in a 
differential manner with respect to M31. The brightest stars in the bulge of 
M31 have $K = 15.6 \pm 0.1$ and so the difference in distance modulus between 
this galaxy and Maffei 1 is $\Delta \mu_0 = 20.0 \pm 0.25 - 15.6 \pm 0.1 = 4.4 
\pm 0.3$. If $\mu_0 = 24.4 \pm 0.1$ for M31 (van den Bergh 2000), A$_V = 5.1 
\pm 0.1$ for Maffei 1 (Buta \& McCall 1983), and $\frac{A_K}{A_V} = 0.11$ 
(Rieke \& Lebofsky 1985), then the unreddened distance modulus of Maffei 1 is 
$28.2 \pm 0.3$, which is in good agreement with the Luppino \& Tonry (1993) 
surface brightness flucuation result ($\mu_0 = 28.1 \pm 0.3$). It is worth 
noting that the stars that dominate surface brightness flucuations in the 
infrared have M$_K = -5.6$ (Luppino \& Tonry 1993); not only are these stars 
significantly fainter than the AGB stars detected in the current 
study, but they are also likely evolving on the first ascent giant 
branch. Thus, the distances estimated here and by Luppino \& Tonry (1993) 
rely on very different stars, and are in this sense independent.

	The RGB-tip, which occurs $\sim 2$ mag in $K$ below the 
AGB-tip in nearby galaxies (e.g. Davidge 2000), provides another means of 
estimating the distance to Maffei 1. The 
results in the current study suggest that the RGB-tip in Maffei 1 should appear 
near $K = 22.0$, which is within the detection threshold of 8 metre 
class telescopes equipped with AO systems. In any event, if Maffei 1 
were as close as 2 Mpc then the RGB-tip would occur at $K = 20$, and this is 
clearly inconsistent with the data presented here; Maffei 1 is therefore too 
distant to significantly affect the dynamics of the Local Group.

	We close by noting that our distance to Maffei 1 is consistent with 
a smooth local Hubble flow just outside the Local Group (Sandage 1986, Ekholm 
et al. 2001). Assuming a smooth Hubble flow with H$_0 = 72 \pm 8$ km sec$^{-1}$ 
Mpc$^{-1}$ (Freedman et al. 2000) and a recession velocity of 
$279 \pm 25$ km sec$^{-1}$ (\S 1) for the IC 342/Maffei group yields a 
distance of $3.9 \pm 0.6$ Mpc, which is in excellent agreement with the 
distance of $4.4^{+0.6}_{-0.5}$ Mpc derived for Maffei 1 from AGB stars.

\vspace{0.5cm}
	It is a pleasure to thank the referee, Dr. Ron Buta, for comments 
that greatly improved the paper.

\clearpage

\begin{center}
FIGURE CAPTIONS
\end{center}

\figcaption
[maffei1figure1.eps]
{The $(K, H-K)$ CMDs of the galaxy and background fields. The locus 
of points with $K < 20$ in each CMD is produced by main sequence stars in 
the Galactic disk. Note that the colors of the main sequence component in 
each field are similar, indicating that the mean reddenings along the two
sight lines are not greatly different. The population of red objects with 
$K > 20.5$ in the galaxy field are identified as stars in Maffei 1.}

\figcaption
[maffei1figure2.eps]
{The $K$ LFs of the galaxy and background fields (top two panels), 
and their differences (lower two panels). n$_{0.5}$ is 
the number of stars per 0.5 mag interval per square 
arcminute. The solid lines in the top two panels are the observed LFs, 
while the dashed lines are the LFs corrected for sample incompleteness. The 
solid lines in the bottom two panels show the difference between the 
completeness-corrected LFs; in the lowermost panel the background field 
LF has been scaled up to match the number of stars with $K < 20$ in the galaxy 
field using the procedure described in \S 3 in an effort to 
correct for field-to-field differences in foreground star counts. 
A statistically significant sample of objects, which 
we identify as stars in Maffei 1, occurs in the galaxy field when $K > 20.5$.}

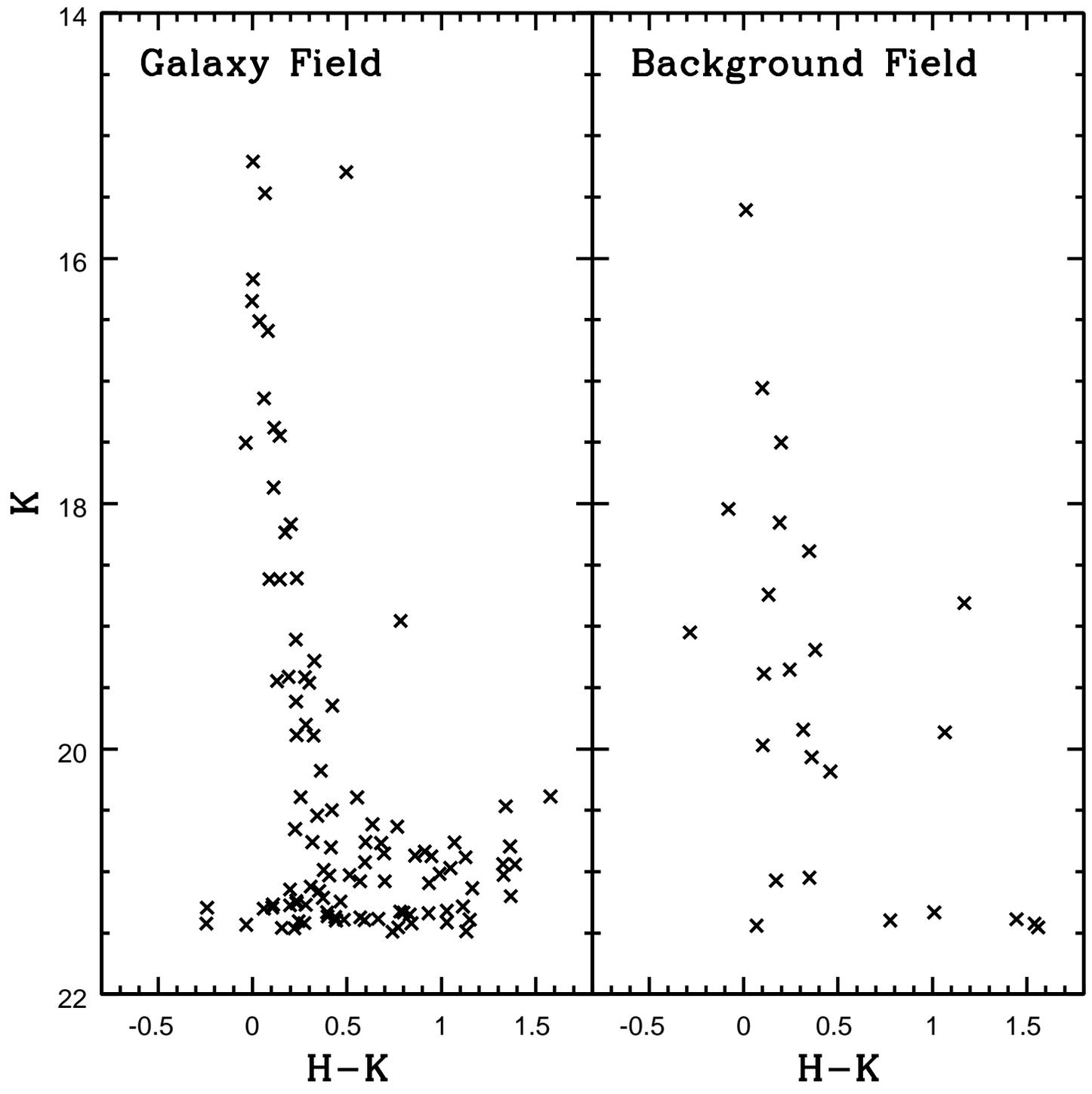
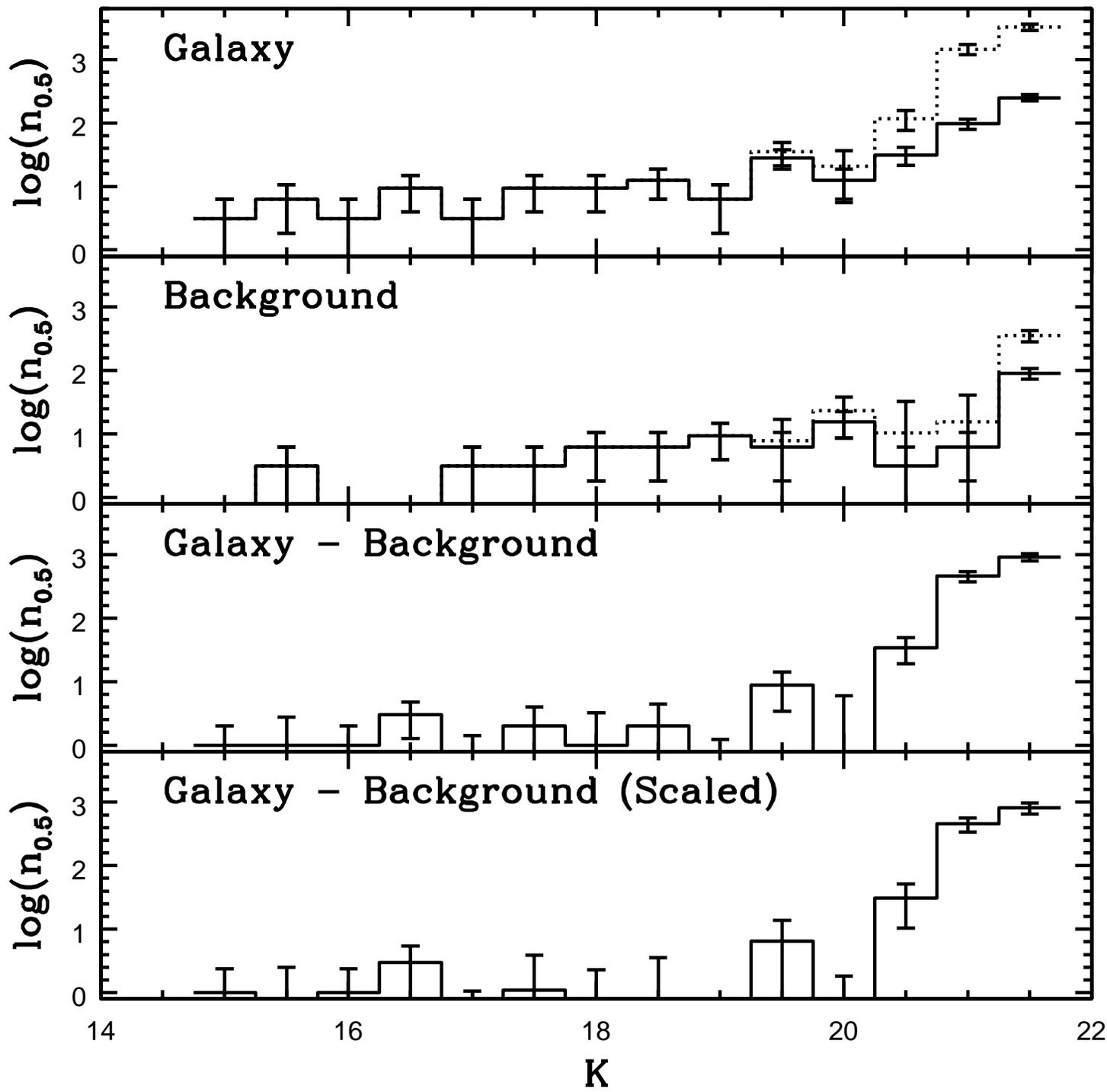
\end{document}